# Composable Generation Strategy Framework Enabled Bidirectional Design on Topological Circuits

Xi Chen, [1] Jinyang Sun, [2] Xiumei Wang, [3][†] Maoxin Chen, [2] Qingyuan Lin, [1] Minggang Xia, [4][*] and Xingping Zhou [5][‡]

[1] *College of Integrated Circuit Science and Engineering, Nanjing University of Posts and Telecommunications, Nanjing 210003, China*

[2] *Portland Institute, Nanjing University of Posts and Telecommunications, Nanjing 210003, China*

[3] *College of Electronic and Optical Engineering, Nanjing University of Posts and Telecommunications, Nanjing 210003, China*

[4] *Department of Applied Physics, School of Physics, Xi'an Jiaotong University, People's Republic of China*

[5] *Institute of Quantum Information and Technology, Nanjing University of Posts and Telecommunications, Nanjing 210003, China*

[†]*wxm@njupt.edu.cn*

[*]*xiamg@mail.xjtu.edu.cn*

[‡]*zxp@njupt.edu.cn*

# Abstract

Topological insulators show important properties, such as topological phase transitions and topological edge states. Although these properties and phenomena can be simulated by well-designed circuits, it is undoubtedly difficult to design such topological circuits due to the complex physical principles and calculations involved. Therefore, achieving a framework that can automatically to complete bidirectional design of topology circuits is very significant. Here, we propose an effective bidirectional collaborative design framework with strong task adaptability, which can automatically generate specific results according to our requirements. In the framework, a composable generation strategy is employed, which involves building a shared multimodal space by bridging alignment in the diffusion process. For simplicity, a series of two-dimensional (2D) Su-Schrieffer-Heeger (SSH) circuits are constructed with different structural parameters. The framework at first is applied to find the relationship between the structural information and topological features. Then, the correctness of the results through experimental measurements can be verified by the automatically generated circuit diagram following the manufacture of Printed Circuit Board (PCB). The framework is demonstrated by achieving good results in the reverse design of circuit structures and forward prediction of topological edge states, reaching an accuracy of 94%. Overall, our research demonstrates the enormous potential of the proposed bidirectional deep learning framework in complex tasks and provides insights for collaborative design tasks.

# I. INTRODUCTION

Topological insulator is an insulator in bulk, but is a conductor on the surface [1]. Inside topological insulators, there are edge states in the electronic band structure, which is different from traditional insulators. Moreover, topological insulators host unique properties, such as topologically protected edge states and bulk–edge separation, which have garnered considerable attention for their role in studying nontrivial band structures and topological protection phenomena, spanning various domains such as electronic solids [2-4], optics [5], microwave metamaterials [6], acoustics [7], and thermoelectric materials [8]. In photonics, topological insulation states are typically achieved by an applied magnetic field [9] or a synthetic gauge field [10]. The edge state of topological protection can be used for unidirectional optical transport in photonics. These fundamental properties and phenomena can be efficiently explored and demonstrated using carefully designed circuits [11-15]. The topological circuit is a sophisticated experimental platform frequently used to simulate topological effects. It allows the precise control of coupling values between nodes by adjusting component values or connections. The approach was illustrated in the previous demonstration where the circuit was comprised of a meticulously arranged network of capacitors and inductors [16].

In recent years, deep learning as a powerful tool for data analysis has been introduced into a number of physical research due to its strong learning and reasoning abilities, ranging from black hole detection [17], gravitational lenses [18], photonic structures design [19], quantum many-body physics, quantum computing, and chemical and material physics [20]. These deep learning frameworks are usually divided into two categories: forward design and reverse design. So far, we have observed significant research achievements in unidirectional designs under various network architectures [21-23], which contribute to establishing a bidirectional "highway" between design parameters and system response [24-25]. However, the inherent bias in traditional unidirectional networks can no longer be evaluated and mitigated by structural information when two different networks are directly connected, which will also lead to the amplification of errors [26-28]. Therefore, it is very important to develop a powerful bidirectional network architecture.

In this work, we propose a universal design method for implementing a two-dimensional (2D) Su-Schrieffer-Heeger (SSH) [15] circuit with the unit cell containing 4 atoms, and verify the correctness of the topological properties by calculating the grounded Laplace matrix. Then, we propose a many-to-many (input and output data come in various forms) framework to realize the bidirectional design of 2D SSH circuits. Our framework consists of three main processes: firstly, different design requirements are uniformly encoded, and then these encoded data are mapped to language-like representations through a projection layer. Secondly, a Large Language Module (LLM) is used as the core to achieve semantic understanding and inference. Thirdly, through the multimodal signal projection of specific instructions, different signals enter different encoders, and then be generated to the circuit structure diagram. In addition, we have increased the input forms of structural parameters, which allows our framework to better understand the intrinsic connections. We create a dataset by varying the circuit structure and parameters. Relying on this dataset, we train our framework successfully. Compared with the existing single forward or reverse design methods, this framework we proposed can perform multiple tasks including circuit topology characterization prediction and reverse topological circuit design at the same time. Besides, more different tasks can be interconnected in real time, thanks to its bidirectionally design ability. This capability is important when dealing with complex bidirectional problems. These problems are commonly encountered in surface structure involvement, such as invisibility cloaks [29], optical computing [30-32], and wireless communication [33].

## II. Two-Dimensional Su-Schrieffer-Heeger Model

In the field of topological physics, the skin effect and the bulk-boundary correspondence of one-dimensional (1D) SSH chain are widely studied [34-40]. A more complex 2D SSH model can be seen as stacking the 1D SSH chains and allowing the strength of the connection between adjacent chains to change. This results in a wide variety of structures for the 2D SSH networks, which typically exhibit a number of unique physical properties. To advance our understanding, we select a model whose unit cell contains 4 atoms as the research object. Therefore, the general Hamiltonian of this 2D SSH model can be uniformly expressed as:

$$H = \sum_{m,n}[(t_x + \delta_x p_{m,n})c^\dagger_{m+1,n}c_{m,n} + (t_y + \delta_y q_{m,n})c^\dagger_{m,n+1}c_{m,n}] + \text{H.c.} \quad (1)$$

where, $c^\dagger_{m,n}$ represents the creation operator for the spineless electron located in $(m,n)$ and $c_{m,n}$ is the annihilation operator for the spineless electron at $(m,n)$. $t_x(t_y)$

represents the hopping integral along the *x(y)-direction*. $\delta_x(\delta_y)$ corresponds to the strength of the lattice-electron coupling along the *x(y)-direction*. $p_{m,n}(q_{m,n})$ reflects the Peierls transition [41] along the *x(y)-direction*. By varying the values of $p_{m,n}$ and $q_{m,n}$, various extended configurations of the 2D SSH model can be obtained. For simplicity, we define $t_x = t_y = t$, $\delta_x = \delta_y = \delta$, $w = t + \delta$ and $v = t - \delta$.

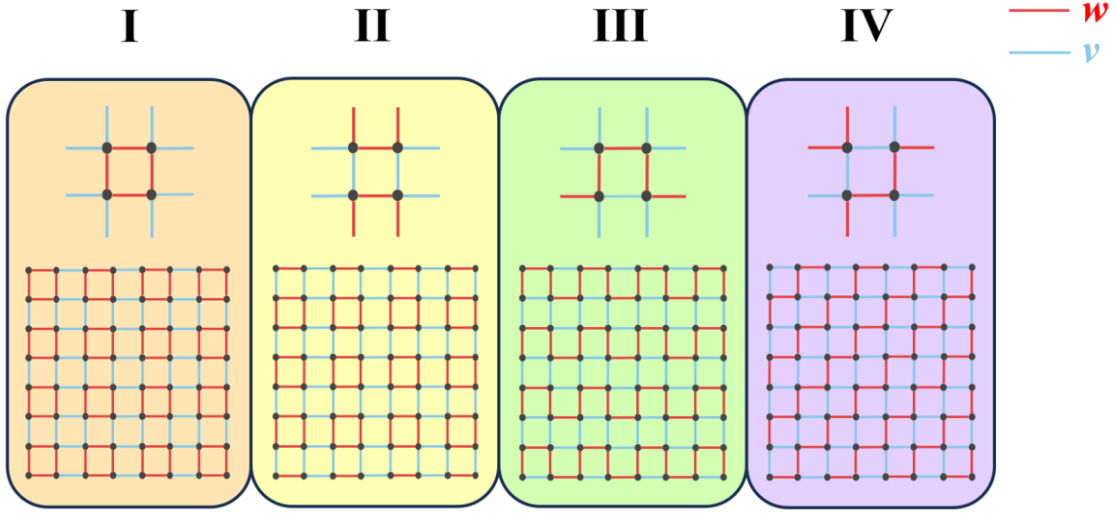

Fig. 1 Four structures of a two-dimensional SSH model with 4-point unit cells. The solid red line represents the jump integral $w$, and the solid blue line represents the jump integral $v$. At the top of each box is the basic periodic unit cell structure.

We plot the schematic diagram for the four possible classes of periodic unit cell structures (limiting the number of atoms in a unit cell to 4 and fixing the connection $w$, $v$) and their open boundary structures in Fig. 1. The Hamiltonian of these four types of structures can be obtained by changing $p_{m,n}(q_{m,n})$. The first type of structure has been studied in many ways, and its Peierls transitions can be described as:

$$p_{m,n} = (-1)^m, q_{m,n} = (-1)^n \qquad (2)$$

Similarly, the Peierls transitions for the next three structures can be formulated as follows:

$$\begin{cases} p_{m,n} = (-1)^{m-1}, q_{m,n} = (-1)^n & \text{(II)} \\ p_{m,n} = (-1)^m, q_{m,n} = (-1)^{m+n} & \text{(III)} \\ p_{m,n} = (-1)^{m+n-1}, q_{m,n} = (-1)^{m+n-1} & \text{(IV)} \end{cases} \qquad (3)$$

Obviously, all 16 different open boundary connection structures can be obtained by swapping the position of jump integral $w$ and $v$. But all of these structures have the symmetrical or rotationally symmetrical relationships with the above four categories. As a result, they also have similar even identical properties. The specific connection methods of all 16 open boundary structures are shown in Appendix A.

For 2D SSH structures, the occurrence of edge states often attracts people's attention and research. These phenomena occur only when periodic boundary conditions are converted to mixed or open boundary conditions. Taking the structure in Fig. 2(a) as an example, we first calculate the band structure under the periodic boundary condition, as shown in Fig. 2(b)-(d). It is clear that there is no separated boundary state in the energy band structure. After converting the periodic boundary condition to the $1\times 8$ mixed boundary condition, the separated boundary states can be identified from its band structure, as shown in Fig. 2(f). In order to visualize the edge state, we then calculate the Hamiltonian of an $8\times 8$ open-boundary model, as shown in Fig. 2(e). Due to the similarity between the open boundary model and the mixed boundary model in terms of band structure, we choose the *A-point* and *B-point* in Fig. 2(f) to observe the edge and bulk effects of the open-boundary model, as shown in Fig. 2(g)-(h).

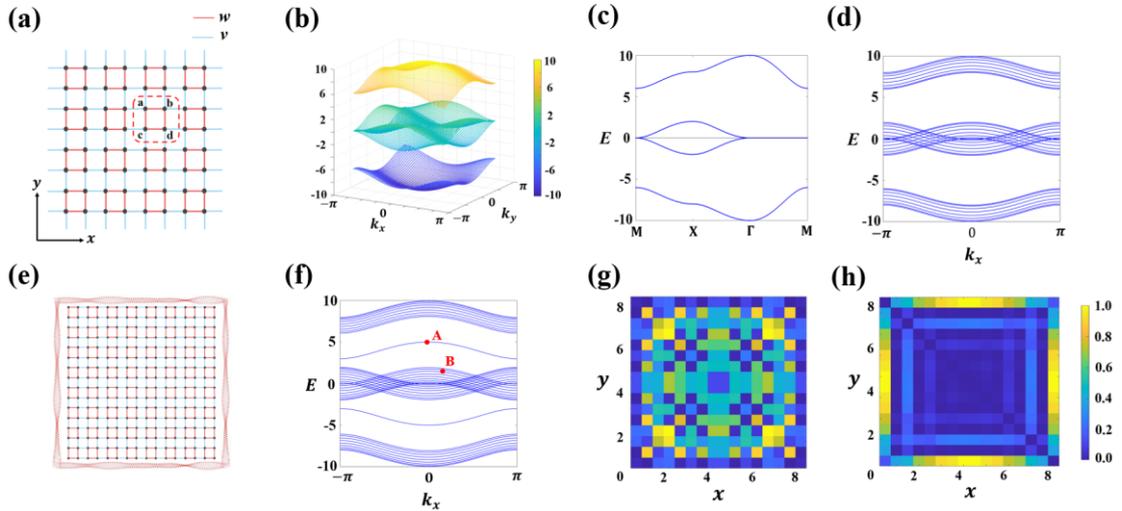

Fig. 2 Band structure and edge effects under different boundary conditions. (a) Schematic diagram of the periodic boundary structure, in which the periodic unit cell is framed by the red dashed box, with $w=4$ and $v=1$. (b) Band structure under periodic conditions. (c) Schematic diagram of the band structure of the first brillouin zone under the periodic conditions. (d) the structure of the $k_x$-related band under the periodic conditions. (e) Schematic diagram of the open boundary structure composed of $8\times 8$ unit cells. (f) Band structure under mixed boundary conditions composed of $1\times 8$ unit cells. (g) Normalized energy intensity distribution which shows the bulk effect of *B-point* in Fig. 2 (f). (h) Normalized energy intensity distribution which shows the edge effect of *A-point* in Fig. 2 (f).

It is evident that the 2D SSH models show various edge states under different conditions. This kind of edge state is worthy for further study and application, but the diversity of topological structures makes the topological edge states unpredictable. Distinguished from the calculation of topological characteristics in traditional theoretical physics, the introduction of circuit model can greatly reduce the difficulty of observing topological characteristics in experiments. Therefore, it is important to combine topology principles with a circuit platform which is practical and flexible.

### III. Two-dimensional Su-Schrieffer-Heeger circuit and its Laplacian matrix

After discussing the 2D SSH theoretical model in **part II**, we take the first structure in Fig. 1(a) as an example to construct the corresponding circuit model, which allows us to observe the topological characteristics more directly. To simulate the band structure, we first consider corresponding the 2D SSH periodic lattice model to an infinite circuit. We also select the periodic lattice model in Fig. 2(a) as an example, and the circuit structure can be easily designed, as shown in Fig. 3(a). And the node voltage vector **V** and the current vector **I** can be correlated by the grounded Laplace matrix **J**. The detailed derivation is shown in Appendix B. Analogous to lattice Hamiltonian, $k_x$ and $k_y$ denote the phases of the block wave vector propagating in the *x* and *y* directions.

We set $L_a = 27\,nH$, $L_b = 220\,nH$ and $C = 1\,nF$. By solving $\det[J(\omega, k_x, k_y)] = 0$, we can easily obtain the band structure under different Bloch oscillating wave vectors $k_x$ and $k_y$, as shown in Fig. 3(c). Obviously, the band structure is very similar to the one obtained in **part I**. Similarly, by solving the ground Laplace matrix of the mixed boundary circuit, we can obtain the band structure, as Fig. 3(d). By observing the energy band structure, we find that the energy bands can be divided into five sections, ranging from 0 MHZ to 70 MHZ. Isolated energy bands appear within the frequency ranges of (31,42) MHZ and (52,60) MHZ, which correspond to the emergence of edge effects. In order to observe the edge state, we build a $6 \times 6$ circuit model, as Fig. 3(b), where the red line indicates the inductance $L_a$, the blue line represents the inductance $L_b$, and the black dot indicates the capacitance C connected to ground. We then use the LTspice simulation software to scan the absorption rate of the bulk nodes (5,4) and edge nodes (8,4) with the range of 0~70 MHZ, as shown in Fig. 3(d).

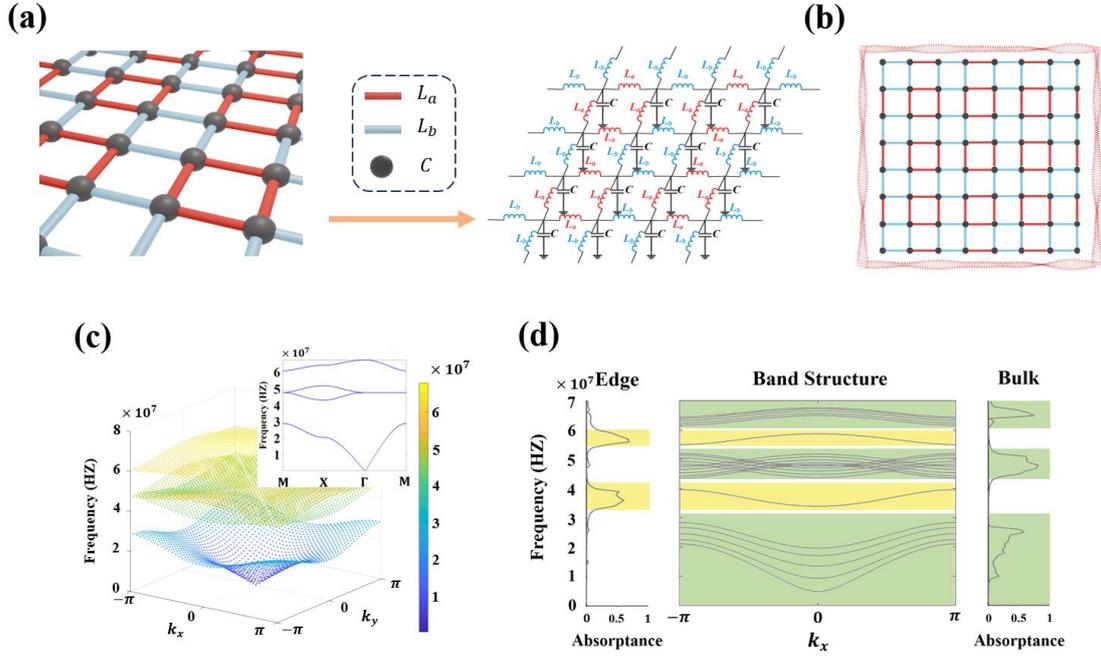

Fig. 3 The circuit structure of a 2D SSH topology circuit and its band structure. (a) The schematic diagram of the 2D SSH atomic structure converted into a circuit structure consisting of capacitors and inductors with different values. The inductance $L_a$ represents the coupling strength between the intracellular capacitors (internal coupling) and the inductance $L_b$ represents the coupling strength between two adjacent cells (external coupling). (b) Structural diagram of the finite 2D SSH circuit. (c) The 2D band structure calculated from the grounded Laplace matrix of the periodic circuit model, whose parameters are selected as: $L_a = 27\ nH$, $L_b = 220\ nH$ and $C = 1\ nF$. (d) The band structure of a finite-size 2D SSH circuit with $1\times 6$ cells along the *x-direction*, and the simulated absorption spectra corresponding to the bulk nodes and edge nodes of the $6\times 6$ 2D SSH circuit. The bulk and edge absorption peaks are highly consistent with the band structure.

Fig. 3(d) shows that the three absorption peaks of the bulk node occur exactly in the frequency range of the bulk band in the band structure diagram (green area in Fig. 3(d)), while the two absorption peaks of the edge node are located between the two bulk bands (yellow area in Fig. 3(d)). The great agreement between the band structure and the absorption rate distribution demonstrates the accuracy of the analysis of topological circuits using the circuit Laplace function. It is not difficult to find that all the bulk states of the finite-size topological circuit can be obtained by the grounded Laplace function of the infinitely periodic circuit model, while the edge states should be calculated from the Laplace function of the finite-size circuit. By varying the values of the capacitor and inductance, we can also easily adjust the size of the bandgap and even change its topological characteristics.

Since the absorption rate of this circuit is not only related to the operating frequency but also to the node position, the circuit can be seen as a position-dependent bandpass filter which can be used to implement directional filtering in finite-size circuits. However, due to the diversity of circuit structures, operating frequencies, and topological characteristics, it is quite difficult to quickly design circuits with specific topological characteristics or predict the topological characteristics of each circuit at different frequencies. Therefore, it is necessary to implement a bidirectional co-design framework.

### IV. Network architecture and training process

Here, we propose a many-to-many coordinated design framework, and use the topological circuit model proposed in **part II** for experiments and tests. This framework adopts a composable generation strategy that involves a shared multimodal space by building aligned bridges during the diffusion process [42]. In this process, data from different modalities will first be mapped to a shared space, thereby achieving the fusion and generation of multimodal data. This method can effectively utilize the complementarity of different modal data, which improves the quality and reliability of the generated results. At the same time, the composability of this strategy also allows it to be easily combined with other deep learning technologies, further improving the performance and flexibility of the system.

As shown in Fig. 4, we depict the framework of our network. The whole operation process can be divided into two parts: training and working. But no matter which process they are, they share a similar set of working module processes: Input Encoding Module ('Encoder' in Fig. 4), LLM-centric Alignment Module ('Input Projection' in Fig. 4), Large Language Understanding Module ('LLM' in Fig. 4), Instruction-following Alignment Module ('Output Projection' in Fig. 4) and Diffusion Decoding Module ('Decoder' and 'Diffusion' in Fig. 4).

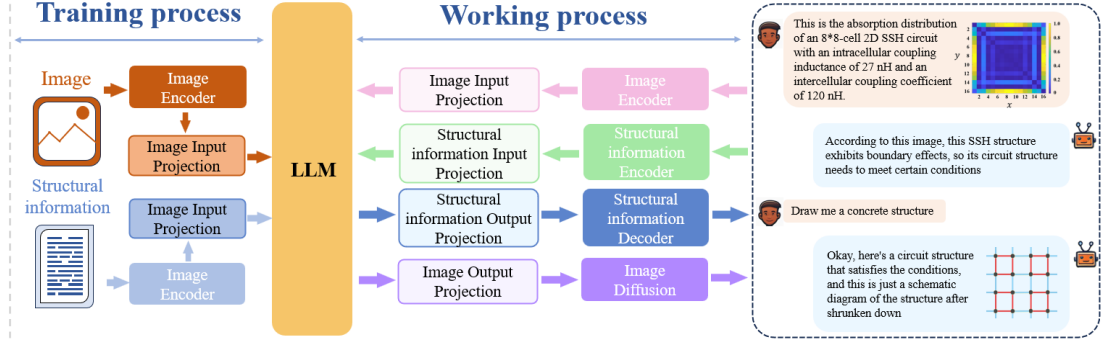

Fig. 4 Network architecture. The training and working process of the network has a similar hierarchy. In the process of data input, data passes through the Input Encoding Module, LLM-centric Alignment Module. For the output of the data, data goes through the LLM Understanding Module, Instruction-following Alignment Module and Diffusion Decoding Module. By connecting the LLM with a multimodal adapter and a diffusion decoder, the intermodal input and output can be realized.

In order to achieve interoperability between modalities, we first encode the input data in the network construction stage. Here, we choose ImageBind, a high-performance encoder submitted by FAIR, Meta AI, to encode the text data and image data into the same feature format [43]. Next, we consider aligning image features with text feature space, which is the main task of the LLM-centric alignment module, because only by aligning text and image features can LLM understand the correspondence between images and text. This is also the key to achieving bidirectional design tasks. Considering that the bidirectional design framework needs to be based on a large amount of data, it is important to correctly match the forward and reverse data, otherwise it is easy to confuse the training data and further lead to the failure to converge. In order to achieve data differentiation and alignment, we classify and pair the existing text and image data, and train the alignment layer based on these paired data, so that it can automatically select and generate corresponding feature labels according to the image features. An example of a concrete text-image pairing is placed in Appendix C.

The LLM-based comprehension and inference module is the core of the entire framework. We adopt Vicuna2 [44], an open-source text-based LLM. The LLM receives inputs from different modules, performs semantic understanding and inference on the inputs, and then directly outputs labeled signal labels and eigenvectors in different states. Since we introduced a pre-trained LLM as the inference core, this also allows our framework to perform a better human-machine collaboration model, which is similar to a conversation between two humans. Compared with the traditional generation mode, we believe that this conversation-like co-design model can be more flexible and convenient to achieve design tasks. After that, the signals and features generated by the LLM are first aligned by the Instruction-following alignment module, which is similar to the LLM-centric alignment module for input process. The main purpose of Instruction-following alignment module is to align the diffusion decoding model with the output instructions of the LLM. In other words, the module receives a multimodal signal with a specific instruction from the LLM and converts the signal labels into the form that can be understood by subsequent multimodal decoders through the output projection layer of the transformer. In this part, we achieve alignment by minimizing the distance between the signal labels of the LLM and the conditional text of the diffusion model. Finally, through the decoder module with different signal labels (Stable diffusion (SD) [45] is used for image synthesis), the network can finally realize the data output of the corresponding modality.

After constructing the network model, we create a dataset according to the topological circuit model given in **part III**, which includes the text data consisting of the parameters, operating frequencies and topological characteristics, and the image data containing the schematic diagram of the circuit structure and the node absorption rate distribution maps. By changing the circuit structure and parameters, we create a dataset containing 8,200 structural parameter data in the form of text and 12,400 absorption rate distribution maps, schematic diagrams of the circuit structure in the form of picture. In order to better understand the interconnections between data, we also improve and refine the text description language. We expand the amount of text by adding a clearer and more detailed description of the circuit model, so as to improve the data utilization efficiency and analysis efficiency of the network model. The expanded text volume is the size of the text data in Appendix C. Then, we train our network model using the dataset we create to make it better adapted to the topological circuit model.

Since our dataset only contains image and text data, we freeze the backbone when training the text encoder, which can greatly reduce the training amount while ensuring the training quality. Therefore, we apply the dataset we built to the training of the overall network, and after about 10,000 rounds of training, the network tends to converge. We then test the trained network.

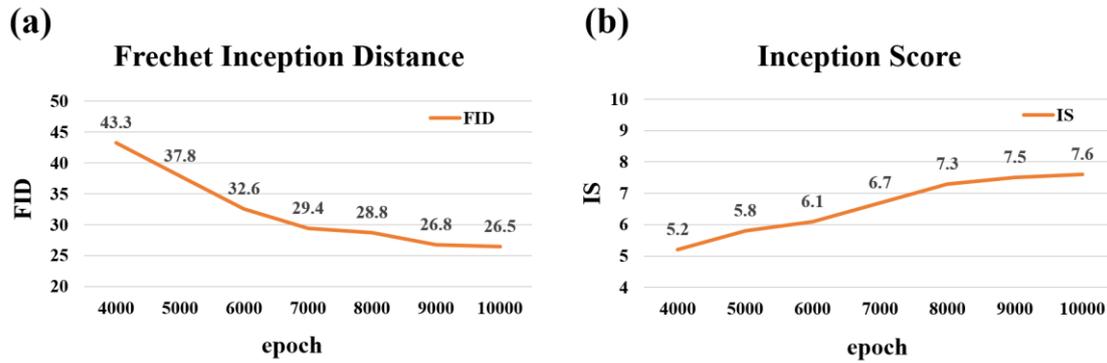

Fig. 5 Change curve of network evaluation index. (a) The FID (↓) trend of the network. (b) The IS (↑) trend of the network. Both indicators are stabilizing at about 10,000 epochs.

We take out the models after 4,000 epoch and calculate the FID (Fréchet Inception Distance) and IS (Inception Score) index in our dataset. As shown in Fig. 5, our model tends to stabilize around 10,000 rounds, with the FID of 26.5 and the IS of 7.6. The specific calculation process of the index is shown in Appendix D. In order to further verify the accuracy of the network, we perform two prediction tasks. The first task is the forward prediction of the circuit absorption rate distribution, as shown in Fig. 6(a). We provide a specific circuit structure (in the form of a picture) and a piece of text with component parameters, operating frequency, and design requirements. While the second task is the reverse design of the circuit structure and operating conditions, as shown in Fig. 6(b). We select a special circuit absorptance distribution diagram that only exhibits edge effects in the *y-direction*, and require our network to give the circuit structure, detailed component parameters and operating frequency which meet the requirements. As shown in Fig. 6, we find that our network is able to give a seemingly correct answer in both tasks, indicating that our work is fairly close to the expected result.

### (a)

This is a schematic diagram of the circuit structure, where the red line segment represents the 14nH inductance, the blue line segment represents the 160nH inductance, and the black dot is the 1nF capacitance, please help me predict the absorption rate indexing of each node at the operating frequency of 62MHZ?

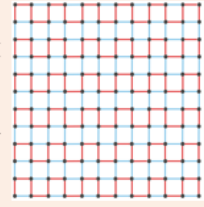

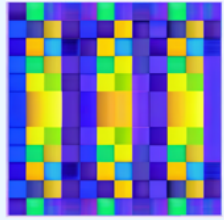
The size of this circuit is 6*6. At a working frequency of 62MHz, the distribution of node absorption rate is shown.

### (b)

This is a 6*6 size circuit node absorption distribution diagram, please use 14nH and 160nH inductors and 1nF capacitors to design a schematic diagram of the circuit structure of the same size and its operating frequency that conform to the diagram.

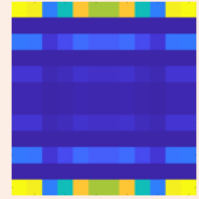

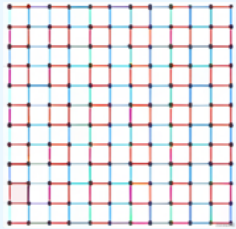
This is a compliant structural design, where the red and blue lines represent 14nH and 160nH inductors, respectively, and the black dots are 1nF capacitors, which operate at a frequency of 47MHZ.
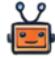

Fig. 6 Schematic diagram of network operation. Since the network is centered on LLM, it can realize human-machine collaborative design tasks in a way that is close to human dialogue. (a) Forward prediction of circuit absorption rate distribution. (b) Reverse design circuit structure and parameters.

## V. Experimental Validation of Topological Properties

In order to verify the correctness of the results given by our network, we design and fabricate two Printed Circuit Boards (PCB) containing 6×6 cells, as shown in Fig. 7(a) and Fig. 7(c). The capacitors are represented by black spheres, and the inductance $L_a$ and inductance $L_b$ are represented by red and blue lines. Their parameter is selected as $L_a = 14\ nH$, $L_b = 160\ nH$ and $C = 1\ nF$, which is the same as the parameter given in Fig. 6. In order to facilitate the measurement of the absorption rate, we extend an SMA connector from each node of the PCB. Based on the operating frequency given by the network, we measure the nodal reflection coefficient $S_{11}$ of the PCB, by a vector network analyzer (Deviser NA7662A) and a 50 Ω coaxial cable. When the input impedance of a node is 50 Ω, the reflection coefficient $S_{11}$ is zero (linear scale), and the absorption rate can be obtained from $1-S_{11}$. As the result, the normalized absorption rate distribution can well represent the energy of different nodes.

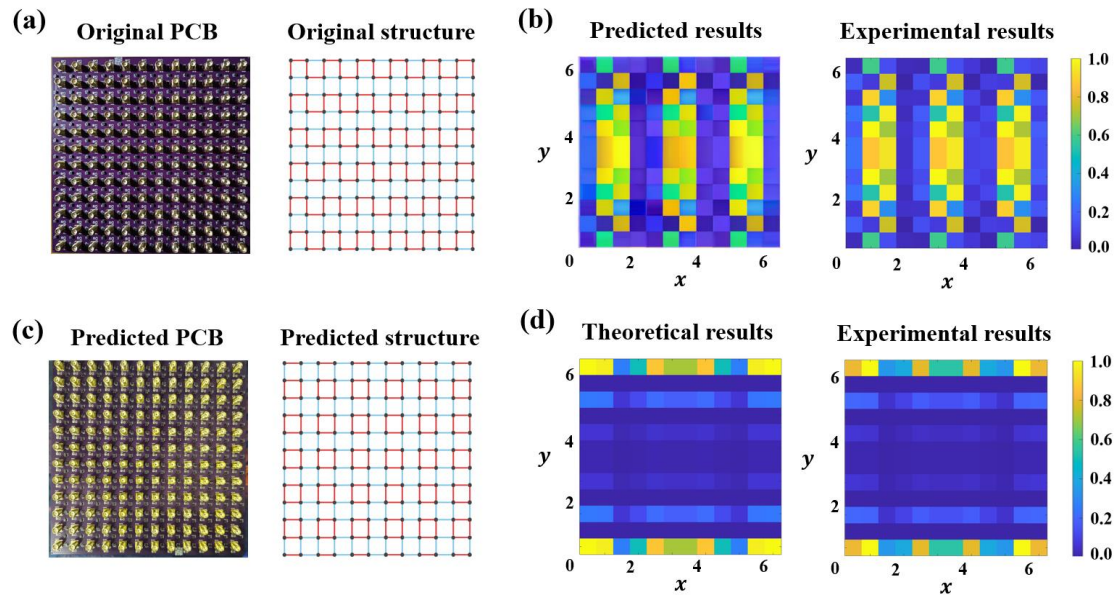

Fig. 7 Fabrication samples, network prediction results and experimental results of the 2D SSH circuits. (a) Photographs and the structural diagram of the 6×6 unit cell samples. We provide this original circuit structure to our network in **part III**, as shown in Fig.6 (a). (c)Photographs and the structural diagram of the 6×6 unit cell samples which is predicted by our network, as shown in Fig. 6 (b). The red and blue lines represent the inductors $L_a$ and $L_b$, the black spots represent the grounded capacitance $C$. (b) The left image shows the experimental measurement results of the circuit structure Fig. 7 (a), and the right image shows the results of the absorption rate distribution predicted by our network. (d) The image on the left shows the experimental measurement results of the circuit structure Fig. 7 (c), and the other one shows the numerical distribution of the node absorption rate calculated by LT spice.

For Fig. 7(a), we plot the normalized absorptivity distribution measured at 62 MHZ operating frequency node by node. As shown in Fig. 7(b), it can be clearly seen that at this operating frequency, the circuit exhibits a special body propagation characteristic in the *y-direction*, which is consistent with the predicted results. In the case of Fig. 7(c), the circuit is automatically designed by the network. According to the structure diagram, we build a physical circuit diagram and measure the absorption rate distribution at the operating frequency of 47 MHZ, as shown in Fig. 7(d). We can clearly see that there is a bright border with two edges, and there is good agreement between the simulation results and the theoretical absorption rate distribution. In addition, we can also intuitively understand the occurrence of such an edge state in one direction. In this model, the unit cell formed by the four nodes connected by the red line only covers all the internal nodes along the *y-direction*, and its edges do not form one unit cell, which is one of the reasons for the formation of edge states. On the contrary, along the *x-direction*, the unit cells cover all internal and edge nodes, and the unit cell of the edges inhibits the formation of edge states, corresponding to a trivial topology.

To further validate the effectiveness of the network, then we select 100 sets of forward and reverse design tasks and verify their accuracy through circuit simulation. After testing, we find that our network shows a high accuracy in both tasks, with an average accuracy rate of 94%. Through the above verification, it can be seen that the bidirectional co-design framework we have built can adapt to the design requirements of topological circuits. It also shows the great potential of the bidirectional co-design framework for more flexible and complex tasks.

## VI. SUMMARY AND OUTLOOK

In our work, we focus on the design of 2D SSH topological circuits with edge states. In order to deeply explore the characteristics of topological circuits and realize the bidirectional automation design of their structures, we propose a multimodal collaborative design framework that takes both forward and reverse directional tasks into account. This framework adopts a composable generation strategy to construct a shared multimodal space by bridging alignment during the diffusion process. We also construct a series of 2D SSH circuits with different parameters and structures. Then, we apply our network to different prediction tasks of these circuits, ultimately achieving good results. For traditional deep learning frameworks, dealing with complex multimodal problems is very difficult due to the singularity of their structure [46]. Notably, the framework we propose provides a solution to such tricky problems.

The framework we propose has great potential to be further extended to other fields. For example, it can provide a versatile and effective solution for areas facing the dilemma of non-uniqueness. Human-machine collaboration is also a major feature of the framework. When we are not satisfied with the current results given by the network, we can modify the framework, instead of starting over. Compared with the traditional imperative generation network, this collaborative generation model, which resembles the form of human communication, is undoubtedly more intelligent and efficient. Going forward, we can further advance the application of multimodal generative framework, such as Sora [47], in the area of physics.

**Acknowledgements**

The authors thank for the support by NUPTSF (Grants No. NY220119, NY221055), National Natural Science Foundation of China (Grant No. 11774278), CMC Rapid Support Project (61409220140). We thank Professor Xiaofei Li and Dr. Dandan Zhu for useful discussions.

# APPENDIX A. SCHEMATIC DIAGRAMS OF ALL THE STRUCTURES OF THE TWO-DIMENSIONAL SU-SCHRIEFFER-HEEGER CIRCUIT

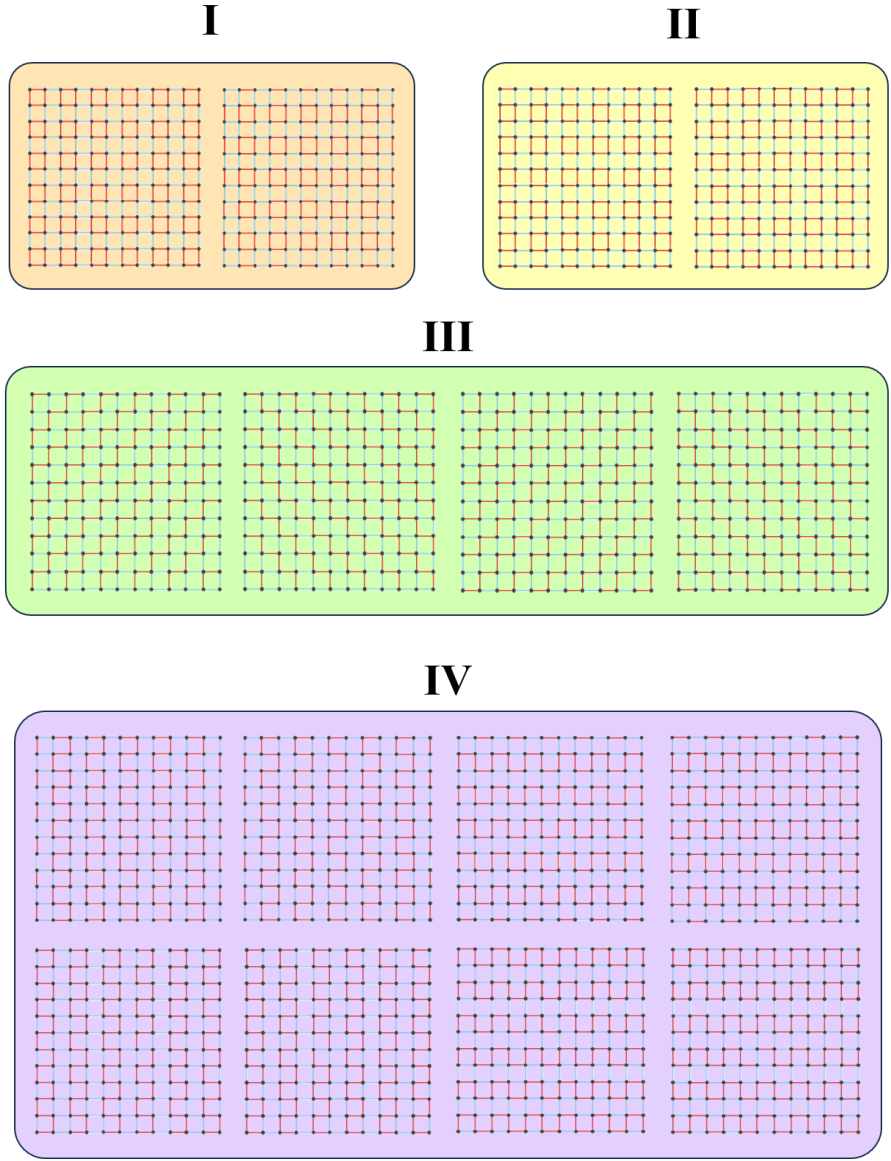

Fig. S1 All 16 two-dimensional Su-Schrieffer-Heeger model configurations with the cell containing 4 atoms. The solid red and blue lines representing jump integrals $w$ and $v$.

As shown in Fig. S1, the all 16 configurations are divided into four groups with different background colors according to Fig. 1 in the article. Structures in each group can actually be obtained by rotation, symmetry, or interchange $w$, $v$. As a result, structures in each group also have similar topological properties.

# APPENDIX B. DERIVATION OF THE LAPLACIAN MATRIX OF THE INFINITE AND FINITE TWO-DIMENSIONAL SU-SCHRIEFFER-HEEGER CIRCUITS

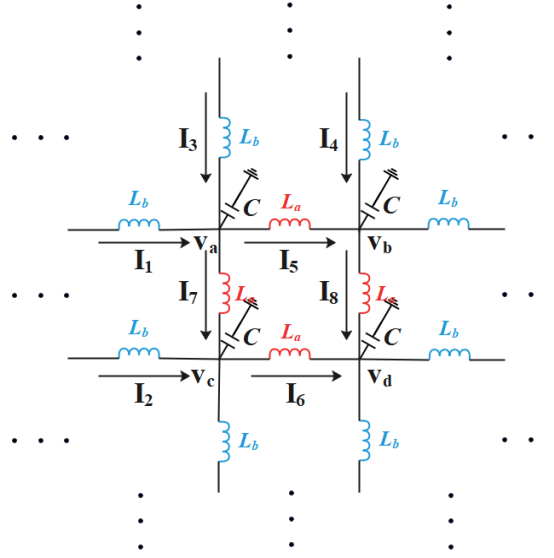

Fig. S2 Schematic diagram of an infinite circuit. The node voltages on the four nodes are $V_a \sim V_d$, and the currents in the corresponding branches are $I_1 \sim I_8$.

Since we are discussing a four-node cell structure circuit in this article, we can study the electrical properties of a single cell under the condition of periodic infinite size. As shown in Fig. S2, we assume voltages $V_a \sim V_d$ on the four grounded capacitors and current directions $I_1 \sim I_8$ in the circuit branches. Applying Kirchhoff's current law to the four nodes, we have:

$$I_1 + I_3 = I_5 + I_7 + \frac{\partial}{\partial t} V_a C \quad , \tag{A1}$$

$$I_5 + I_4 = I_1 e^{ik_x} + I_8 + \frac{\partial}{\partial t} V_b C \quad , \tag{A2}$$

$$I_2 + I_7 = I_6 + I_3 e^{ik_y} + \frac{\partial}{\partial t} V_c C \quad , \tag{A3}$$

$$I_6 + I_8 = I_2 e^{ik_x} + I_4 e^{ik_y} + \frac{\partial}{\partial t} V_d C . \tag{A4}$$

Based on the direction of the current given in Fig. S2, we can write the current flowing out of each node as:

$$I_a = \frac{-\left(V_b e^{-ik_x} - V_a\right)}{i\omega L_a} + \frac{\left(V_a - V_b\right)}{i\omega L_b} - \frac{\left(V_c e^{-ik_y} - V_a\right)}{i\omega L_a} + \frac{\left(V_a - V_c\right)}{i\omega L_b} + V_a i\omega C, \quad (A5)$$

$$I_b = \frac{-\left(V_a - V_b\right)}{i\omega L_b} + \frac{\left(V_b - V_a e^{-ik_x}\right)}{i\omega L_a} - \frac{\left(V_d e^{-ik_y} - V_b\right)}{i\omega L_a} + \frac{\left(V_b - V_d\right)}{i\omega L_b} + V_b i\omega C, \quad (A6)$$

$$I_c = \frac{-\left(V_d e^{-ik_x} - V_c\right)}{i\omega L_a} + \frac{\left(V_c - V_d\right)}{i\omega L_b} - \frac{\left(V_a - V_c\right)}{i\omega L_b} + \frac{\left(V_c - V_a e^{ik_y}\right)}{i\omega L_a} + V_c i\omega C, \quad (A7)$$

$$I_d = \frac{-\left(V_c - V_d\right)}{i\omega L_b} + \frac{\left(V_d - V_c e^{ik_x}\right)}{i\omega L_a} - \frac{\left(V_b - V_d\right)}{i\omega L_b} + \frac{\left(V_d - V_b e^{ik_y}\right)}{i\omega L_a} + V_d i\omega C. \quad (A8)$$

Then we can write the eq. (A5) ~ eq. (A8) as follows:

$$\begin{pmatrix} I_a \\ I_b \\ I_c \\ I_d \end{pmatrix} = J \begin{pmatrix} V_a \\ V_b \\ V_c \\ V_d \end{pmatrix}, \quad (A9)$$

where $J$ is the grounded Laplace matrix as follows:

$$J = i\omega \begin{pmatrix} C - \frac{2}{\omega^2 L_a} - \frac{2}{\omega^2 L_b} & \frac{e^{-ik_x}}{\omega^2 L_a} + \frac{1}{\omega^2 L_b} & \frac{e^{-ik_y}}{\omega^2 L_a} + \frac{1}{\omega^2 L_b} & 0 \\ \frac{e^{ik_x}}{\omega^2 L_a} + \frac{1}{\omega^2 L_b} & C - \frac{2}{\omega^2 L_a} - \frac{2}{\omega^2 L_b} & 0 & \frac{e^{-ik_y}}{\omega^2 L_a} + \frac{1}{\omega^2 L_b} \\ \frac{e^{ik_y}}{\omega^2 L_a} + \frac{1}{\omega^2 L_b} & 0 & C - \frac{2}{\omega^2 L_a} - \frac{2}{\omega^2 L_b} & \frac{e^{-ik_x}}{\omega^2 L_a} + \frac{1}{\omega^2 L_b} \\ 0 & \frac{e^{ik_y}}{\omega^2 L_a} + \frac{1}{\omega^2 L_b} & \frac{e^{ik_x}}{\omega^2 L_a} + \frac{1}{\omega^2 L_b} & C - \frac{2}{\omega^2 L_a} - \frac{2}{\omega^2 L_b} \end{pmatrix}. \quad (A10)$$

Then, by solving $\det[J(\omega, k_x, k_y)] = 0$, we can easily obtain the corresponding band structure in the case of different Bloch oscillation wave vectors. In the same way, the electrical properties of a finite-size circuit can still be solved by its ground Laplace matrix [S1],

$$\mathbf{I} = J\mathbf{V}, \quad (A11)$$

where V and I are the voltage and current vectors composed of node voltage and outflow node current. The Laplace grounded matrix $J$ can be expressed as,

$$J = D - C + W, \quad (A12)$$

where, $W$ and $D$ are diagonal matrices that contain the total conductance from each node to ground and to the rest of the circuit, respectively. $C$ is the adjacency matrix of the circuit diagram, and its edges are weighted by its conductance, a simple example is given in Fig. S3.

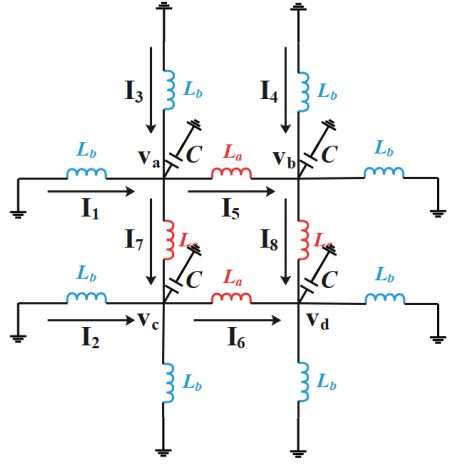

Fig. S3 Schematic diagram of finite circuit structure. Four nodes are grounded through capacitor $C$ and inductor $L_b$ respectively.

According to the circuit structure of Fig. S3, it is easy to write out the components of the Laplace matrix, as follows.

Circuit grounded matrix:

$$W = \begin{pmatrix} 2(i\omega L_b)^{-1} + i\omega C & 0 & 0 & 0 \\ 0 & 2(i\omega L_b)^{-1} + i\omega C & 0 & 0 \\ 0 & 0 & 2(i\omega L_b)^{-1} + i\omega C & 0 \\ 0 & 0 & 0 & 2(i\omega L_b)^{-1} + i\omega C \end{pmatrix}. \quad (A13)$$

Adjacency matrix:

$$D = \begin{pmatrix} 2(i\omega L_a)^{-1} & 0 & 0 & 0 \\ 0 & 2(i\omega L_a)^{-1} & 0 & 0 \\ 0 & 0 & 2(i\omega L_a)^{-1} & 0 \\ 0 & 0 & 0 & 2(i\omega L_a)^{-1} \end{pmatrix}. \quad (A14)$$

Total node conductance matrix:

$$C = \begin{pmatrix} 0 & (i\omega L_a)^{-1} & (i\omega L_a)^{-1} & 0 \\ (i\omega L_a)^{-1} & 0 & 0 & (i\omega L_a)^{-1} \\ (i\omega L_a)^{-1} & 0 & 0 & (i\omega L_a)^{-1} \\ 0 & (i\omega L_a)^{-1} & (i\omega L_a)^{-1} & 0 \end{pmatrix}. \quad (A15)$$

# APPENDIX C. AN EXAMPLE OF THE MATCHING ABSORPTION, STRUCTURAL PARAMETERS AND CIRCUIT STRUCTURE

**Absorption distribution (Image)**

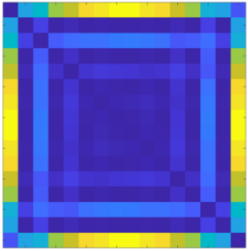

**Structural parameters (Text)**

This is an absorptivity distribution diagram, where the brighter the color indicates the higher the absorptivity at that location, it is from a finite 2D SSH circuit model of a specific structure, the structure is as follows, the size is selected as 8*8, the red line represents the capacitance La, and its value is selected as 27nH. The value is set to 220nH. The black dot represents the ground capacitance, and the capacitance value is 1nF. The circuit works at a frequency of 52MHZ, and it can be clearly seen that the distribution of the node absorption rate shows obvious boundary effects, and similar topological boundary effects have more important significance in circuit design, and this kind of topological circuit may play a special role in specific situations.

**Circuit structure (Image)**

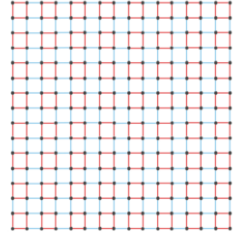

Fig. S4 Schematic diagram of text-image matching data.

In order to achieve efficient text alignment and mining, we improve the description language of the dataset. We expand the amount of text by adding a clearer and more detailed description of the circuit model and its internal logical relationships, so that the network model can improve the efficiency of data utilization and analysis. In the training process, such sets of data are randomly combined and fed, which can not only improve the efficiency of data use, but also promote the efficiency of the network model to mine the internal relationship of data.

# APPENDIX D. THE CALCULATION FORMULA FOR *FID SCORE* AND *IS SCORE*

*1. FID score:* *FID score* is a metric used to evaluate the quality of generated images [S2]. It is specifically used to evaluate the performance of images generated by the model. Its calculation formula is as follows:

$$FID(P,Q) = \|u_P - u_Q\|^2 + Tr(C_p + C_Q - 2\sqrt{C_p C_Q}), \quad (A16)$$

where $P$ represents the distribution of theoretical image, $Q$ represents the distribution of the generated image, $u_P$ and $u_Q$ represents the eigenvectors of images $P$ and $Q$ respectively, $C_P$ and $C_Q$ represents the covariance matrix of the eigenvectors of two distributions respectively, $Tr(...)$ denotes the trace operation of the matrix, and $\|\cdots\|^2$ denotes the square of the Euclide. *FID score* can evaluate the difference between the generated model and the real data distribution. The lower the value, the closer the generated image is to the real image.

*2. IS score:* *IS score* is an evaluation metric in the form of convolutions based on the design of the generative model after training [S3]. The input is a picture tensor, and the output is a probability distribution in the form of a high-dimensional vector. By calculating the distance and divergence between these elements, IS score can measure the sharpness and diversity of the image. The higher the IS score, the better. The detailed calculation formula is as follows:

$$\text{IS}(G) = \exp\left(\frac{1}{N}\sum_{i=1}^{N} D_{KL}(p(y|x) \| \overline{p}(y))\right), \quad (A17)$$

where $G$ is the generative model we use, $p(y|x)$ is the probability vector distribution generated by a given input image $x$, which is the average probability vector of all input images. $D_{KL}$ is the Kullback-Leibler divergence for $p(y|x)$ and $\overline{p}(y)$, and the formula is eq. (A18).

$$D_{KL}(P\|Q) = \sum_i P(i)\log\frac{P(i)}{Q(i)}. \quad (A18)$$

Therefore, we can evaluate the quality of the images given by the network and the diversity of the generated images by the *IS score*.

# APPENDIX E. PHOTOGRAPHS OF EXPERIMENTAL INSTRUMENTS AND CIRCUIT BOARDS

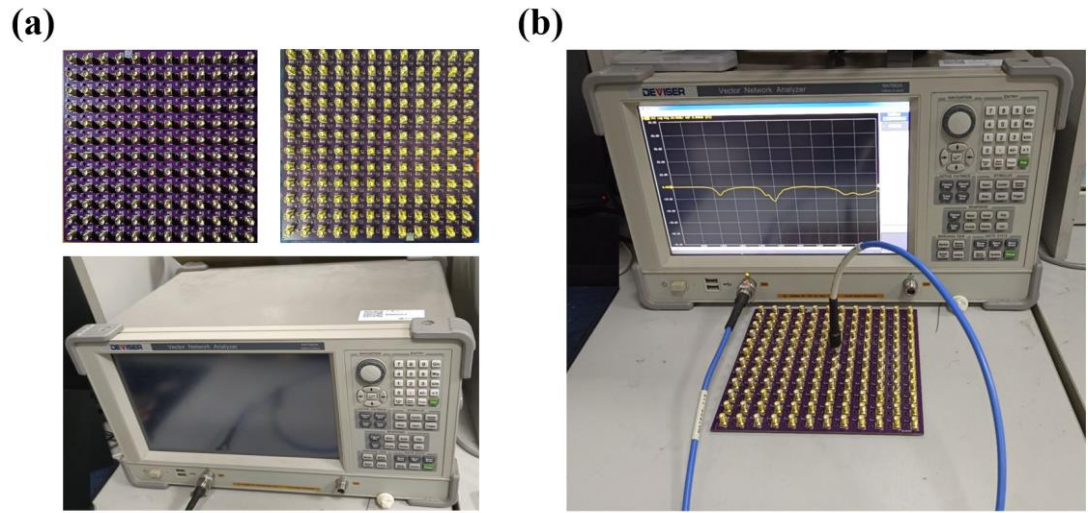

Fig. S4 The photos of experimental instrument and circuit boards. (a) PCB & Vector Network Analyzer (Deviser NA7662A). (b) Photographs of the measurement operation of the experimental instrument.